\documentclass[prd,twocolumn,showpacs,amsmath,amssymb]{revtex4}
\usepackage{amsfonts}

\usepackage{amssymb}
\usepackage{mathrsfs}
\usepackage{txfonts}

\usepackage{graphicx}
\usepackage{dcolumn}
\usepackage{bm}

\begin{document}

\title{Loop quantum cosmology: The horizon problem and the probability of inflation}

\author{Long Chen}
\email{chen_long@mail.bnu.edu.cn}
\affiliation{Department of Physics, Beijing Normal University, Beijing 100875, China}
\author{Jian-Yang Zhu}
\thanks{Corresponding author}
\email{zhujy@bnu.edu.cn}
\affiliation{Department of Physics, Beijing Normal University, Beijing 100875, China}
\date{\today}
\begin{abstract}
Anomaly-free perturbations of loop quantum cosmology reveal a deformed space-time structure, in which the signature changes when the energy density is $\rho=\rho_c/2$. Furthermore, in loop quantum cosmology, one can obtain an effective causal structure only for a low density region ($\rho\leq\rho_c/2$), which gives a natural initial condition to consider the horizon problem. Choosing the initial value at $\rho(0)=\rho_c/2$ in this paper, we investigate the horizon problem and the probability of inflation in the framework of loop quantum cosmology. Two models are considered: the quadratic inflation and the natural inflation. We use the Liouville measure to calculate the probability of inflation which solves the horizon problem, and find that, for the quadratic inflation model, the probability is very close to unity, while for the natural inflation model, the probability is about $35\%$.
\end{abstract}
\maketitle

\section{Introduction}
Inflation, as a necessary supplement to the standard cosmological model, can solve many long-standing problems such as the horizon problem, the flatness problem, etc. Inflation models can also provide a natural explanation of the structure formation. However, whether the inflation itself is probable is also an important question which many authors have considered in Refs. \cite{Gibbons1,Belinsky,Hollands,Kofman,Hartle,Gibbons2,Corichi,Remmen}. For considering this question, two problems should be addressed: one is to find a starting point to count $e$-foldings, the other one is to define a measure to calculate the probability of inflation solutions which give enough $e$-foldings. The former one is a problem because in classical cosmology, due to the existence of the initial singularity, there is no clear starting point to begin one's counting of the $e$-foldings. Since general relativity (GR) is not credible at high densities and curvatures, one may take the starting point at Planck scale which is reasonable but not clear. For the second problem, the Liouville measure, established by Gibbons, Hawking, and Stewart \cite{Gibbons1}, can be used as a candidate measure to calculate the probability. However, in flat, homogeneous, isotropic models, the total Liouville measure of the space of solutions is infinite; thus, one needs a regularization scheme. As a result, the obtained measure depends on the choice of energy density when the regularization is introduced \cite{Corichi2011}. In \cite{Gibbons2}, such choice was taken at the end of inflation when $|\Omega_k|\sim 1$, while in \cite{Belinsky,Remmen}, the authors choose the constant density surface at Planck scale. Their results are quite different: the former gives a quite low probability if inflation can produce enough $e$-foldings, while the latter gives almost one probability for inflation.

In loop quantum cosmology (LQC), big bang singularity is resolved \cite{Ashtekar2006,Vandersloot}. One can start counting the number of e-folding and define a probability distribution at the bounce \cite{Ashtekar,Ashtekar2} or the remote past before the bounce \cite{Linsefors}, where both methods support inflation in the quadratic inflation model. However, because in a bouncing world, any particles have infinite time to have chances to interact with others, one may puzzle whether the horizon problem, which inflation theories try to address, exists. To make this question clear, we need a space-time structure of loop cosmology which has been derived from anomaly-free perturbations of loop quantum cosmology recently \cite{Cailleteau2012a,Cailleteau2012b,Cailleteau2014}. Unlike in the standard cosmological model, where the space-time structure is inserted in the space-time metric, the effective space-time structure of loop cosmology is concealed in the homogeneous model because we obtain the theory by quantizing the symmetry reduced system, and we still miss the full theory of quantum gravity even in some effective level. However, because the constraint algebra needs to be anomaly-free for a first-class constraint system, and by using the theory of anomaly-free perturbations, one fortunately finds the hidden causal structure. As a result, the modified space-time structure has an unexpected property which may address the above puzzling: signature changes at a critical density, $\rho=\rho_c/2$. Furthermore, with the help of perturbation equations of gravity and matter, we can define the effective causal structure by the characteristic, from which we know no causal structure exists at high density ($\rho>\rho_c/2$).

For the above reasons, different from Refs.\cite{Ashtekar,Ashtekar2,Linsefors}, our view is to choose the initial value at $\rho=\rho_c/2$. With the effective causal structure in loop quantum cosmology, we will consider the horizon problem and the probability of inflation in the framework of loop quantum cosmology.

This paper is organized as follows: in Sec.II, we present some results of loop quantum cosmology and its anomaly-free perturbations, and then introduce the effective causal structure of loop cosmology which our construction is based on. In Sec.III, we use the effective causal structure to consider the horizon problem. In Sec.IV, the measure is derived from the Liouville measure of a canonical system. In Sec.V and Sec.VI, we consider the quadratic inflation model and natural inflation, respectively. In Sec.VII some conclusions are given.

\section{\label{sec:level2} Results of Loop quantum cosmology and its perturbations}
In this section, we first present some basic results for a homogeneous model, and then consider its perturbations.
In the spatially flat isotropic model of LQC, one has to first introduce an elementary cell $\cal{V}$ and restrict all integrations to this cell. One can fix a fiducial flat metric ${}^oq_{ab}$ and denote the volume of the elementary cell $\cal {V}$ by $V_0$ in this geometry. The gravitational phase space variables are the connections $A^i_a$ and the density-weighted triads $E^a_i$, which can be expressed as
\begin{equation}\label{1}
A^i_a=cV_0^{-1/3}{}^0\omega^i_a ~~\text{and}~~ E^a_i=pV_0^{-2/3}\sqrt{{}^0q}\;{}^0e^a_i,
\end{equation}
where $\left({}^0\omega^i_a,{}^0e^a_i\right)$ are a set of orthogonal co-triads and triads compatible with ${}^oq_{ab}$ and adapted to the edges of the elementary cell $\mathcal{V}$. The basic Poisson bracket is given by $\{c,p\}=\kappa \gamma/3,$
where $\kappa=8 \pi G$, $G$ is the Newton's constant and $\gamma$ is the Barbero-Immirzi parameter. The matter phase space variables are $\phi$ and $p_{\phi}$, whose Poisson bracket is $\{\phi,p_{\phi}\}=1$.

LQC generates two main classes of effective corrections to the constraints, called the inverse-volume corrections and the holonomy corrections. In this paper, we focus on the holonomy corrections. In the $\bar{\mu}$-scheme of holonomy corrections \cite{Ashtekar2006}, we have the effective constraint:
\begin{equation}
  C_{\text{eff}}=-\frac{3}{\kappa\gamma^2}\left(\frac{\sin(\bar\mu c)}{\bar\mu}\right)^2\sqrt{p}+\frac{p_\phi^2}{2p^{3/2}}+p^{3/2}V(\phi),
\end{equation}
where $V(\phi)$ is the potential of the scalar field, $\bar\mu=\sqrt{\Delta /p}$, and $\Delta$ relates to the minimum nonzero eigenvalue of the area operator from the full theory (LQG).
For convenience, we use the following variables: $b:=\bar\mu c$ and the volume of the elementary cell ${\cal{V}}=p^{3/2}$ whose Poisson bracket is $\{b,{\cal{V}}\}=4\pi G\gamma \sqrt{\Delta}$. In these variables, the constraint becomes:
\begin{equation}\label{constraint}
 C_{\text{eff}}=-\frac{3}{\kappa\gamma^2\Delta}{\cal{V}}\sin^2(b)+\frac{p_\phi^2}{2{\cal{V}}}+{\cal{V}}V(\phi).
\end{equation}
Using the constraint and the equation of $\cal{V}$ and $\phi$, one can get a modified Friedmann equation as the following form
\begin{equation}\label{Friedmann eqn}
  H^2=\frac{8\pi G}{3}\rho\left(1-\frac{\rho}{\rho_c}\right),
\end{equation}
where $H$ is the Hubble factor, $\rho$ the energy density of the matter content, expressed as $\rho=\frac{1}{2}\dot\phi^2+V(\phi)$ and $\rho_c:=\frac{3}{\kappa \gamma^2 \Delta}\simeq0.41\rho_P$ \cite{Ashtekar2006}.

The equation of motion of $\phi$ takes the standard form:
\begin{equation}\label{matter's eqn}
  \ddot\phi+3H\dot\phi+\frac{dV}{d\phi}=0.
\end{equation}
Since $b$ decreases with time: $\frac{db}{dt}=-4\pi G\gamma\sqrt{\Delta} \dot{\phi}^2<0$, one can view $b$ as an internal time which will be important in defining the measure in Sec.IV.

The perturbations of background is needed even for a homogenous model because the causal structure of the effective space-time is not apparent in the homogenous part while its perturbation equations contain the key elements. Anomaly-free perturbations with holonomy corrections have been derived in the series of papers, Refs.\cite{Cailleteau2012a,Cailleteau2012b,Cailleteau2014}. We only present their conclusions, and then get the causal structure which will be useful in our consideration.

Loop quantum cosmology is the symmetry reduced version of loop quantum gravity, and it utilizes key elements of full theory. But when one inserts linear perturbations, anomalies appear and the modified constraints do not form a closed algebra. For eliminating these anomalous terms, one needs to add some ``counterterms". As a result, the constraint algebra is deformed, which can be seen from the Poisson bracket of two scalar constraints
\begin{equation}
\left\{ H[N_1],H[N_2]\right\} =\Omega D\left[ \frac{\bar{N}}p\partial
^a\left( \delta N_2-\delta N_1\right) \right] ,  \label{constraint algebra}
\end{equation}
where $\Omega=\cos(2b)=1-2\rho/\rho_c$. In general relativity, as we know, the factor $\Omega\equiv 1$; thus, LQC deforms the constraint algebra. A surprising thing is at high density (when $\rho > \rho_c/2$ ), $\Omega<0$, which means space-time is more like Euclidean space where the factor $\Omega$ in the Poisson bracket between two scalar constraints is $-1$ \cite{Cailleteau2012a}. The deformed space-time structure also affects the perturbations equations ( in conformal time $\eta$ )
\begin{equation}\label{per-eqn}
  v_{S(T)}''-\Omega\nabla^2v_{S(T)}-\frac{z''_{S(T)}}{z_{S(T)}}v_{S(T)}=0,
\end{equation}
where $v_{S(T)}$ are the scalar or tensor Mukhanov-Sasaki variables.
When $\Omega<0$, the equations become elliptic would make the perturbations instable if one considers these problems as initial-value problems, which will destroy the background dynamics. The argument for the instability of the initial-value problem can be seen in Ref.\cite{Bojowald}, and a strict proof of the instability of a special elliptic equation can be seen in Ref.\cite{Lieberstein}. For understanding this problem, the authors of Ref.\cite{Bojowald} proposed a mixed-type characteristic problem recently while Mielczarek \cite{Mielczarek2014a,Mielczarek2012,Mielczarek2014b} supposed a possible second order phase transition happening at $\rho=\rho_c/2$.
Note that the authors in Ref.\cite{Mielczarek2014c} have considered the tensor power spectrum of cosmological perturbations with initial conditions imposed around $\rho=\rho_c/2$, where they find the cubic shape of the initial power spectrum at $\rho=\rho_c/2$ is favored.
In this paper, we only consider the background dynamics, but for the consistence with perturbations, we think taking the expanding phase at $\rho=\rho_c/2$ as the initial time is appropriate even for background dynamics. Furthermore, that the perturbation equations having characteristics gives an effective causal structure only when $\rho<\rho_c/2$, which has be used in Refs.\cite{Bojowald,Mielczarek2012}:
in conformal time $\eta$, the propagating velocity $v\equiv \frac{dx}{d\eta}$ of any information should be less than $\sqrt{\Omega}$ which agrees with classical theory when $\rho\ll \rho_c$. For later convenience, we will use the proper time; thus, the propagating velocity satisfies
\begin{equation}\label{causal structure}
  \frac{dx}{dt}\leqslant \frac{\sqrt{\Omega}}{a}.
\end{equation}
Using this formula, one can define the particle horizon
\begin{equation}
d_H(t):=a(t)\int^t_{0}\frac{\sqrt{\Omega}}{a(t')}dt',
\end{equation}
where time 0 denotes the initial time with $\rho=\rho_c/2$ and the angular-diameter distance
\begin{equation}
d_A(t):=a(t)\int_t^{t_0}\frac{\sqrt{\Omega}}{a(t')}dt',
\end{equation}
where $t_0$ means today.

\section{\label{sec:level3} The horizon problem in loop quantum cosmology }
One can see that if we use the horizon of LQC defined in the last section, the horizon problem of cosmology in the standard model becomes more severe. As we know, in the standard model of cosmology, the relation of density with the scale factor is
\begin{equation}\label{density eqn}
  \frac{\rho}{\rho_0}=\Omega_\Lambda+\Omega_M\left(\frac{a_0}{a}\right)^3+\Omega_R\left(\frac{a_0}{a}\right)^4,
\end{equation}
where the subscript ``0" denotes the current value,  $\rho_0:=\frac{8\pi G}{3H_0^2}$, and from observation \cite{Planck}, $\Omega_\Lambda\simeq0.685$, $\Omega_M\simeq0.315$, $\Omega_R\simeq 9.2\times10^{-5}$ and $H_0=67.31 \text{km}\cdot \text{s}^{-1}\cdot \text{Mpc}^{-1}$. In loop quantum cosmology, because the equation of matter is the same with classical theory [see Eq.(\ref{matter's eqn})], we still suppose the relation in Eq.(\ref{density eqn}). By using the modified Fridemann equation in Eq.(\ref{Friedmann eqn}), we can find that the ratio between the particle horizon and the angular-diameter distance in LQC is smaller than general relativity
\begin{eqnarray}
\nonumber \frac{d_H(t)}{d_A(t)}&=& \frac{\int^{t}_{0}\frac{\sqrt{\Omega}}{a(t)}dt}{\int^{t_0}_{t}\frac{\sqrt{\Omega}}{a(t)}dt}
=\frac{\int^{a}_{a(t=0)}\sqrt{\frac{1-2\frac{\rho}{\rho_c}}{1-\frac{\rho}{\rho_c}}}\frac{da}{a^2\sqrt{\frac{\kappa}{3}\rho}}}
{\int^{a_0}_{a}\sqrt{\frac{1-2\frac{\rho}{\rho_c}}{1-\frac{\rho}{\rho_c}}}\frac{da}{a^2\sqrt{\frac{\kappa}{3}\rho}}} \\
&<& \frac{\int^{a}_{0}\frac{da}{a^2\sqrt{\frac{\kappa}{3}\rho}}}{\int^{a_0}_{a}\frac{da}{a^2\sqrt{\frac{\kappa}{3}\rho}}}=\frac{d_H^{GR}(t)}{d_A^{GR}(t)},
\end{eqnarray}
which means the horizon problem is more severe than GR. Inflation theory tries to address the horizon problem by inserting a scalar field to drive an almost exponential expansion before transforming into a hot universe dominated by radiation. The horizon problem is solved when $d_H(t_1)>d_A(t_1)$, where $t_1$ denotes the time of the end of inflation. If we suppose the reheating process did not produce many $e$-foldings, then the angular-diameter distance $d_A$ at $t_1$ can be approximated by\cite{Weinberg} as
\begin{eqnarray}
  \nonumber d_A(t_1) &=& a_1\int_{t_1}^{t_0}\frac{\sqrt{\Omega}dt}{a}\simeq a_1\int_{t_1}^{t_0}\frac{dt}{a} \\
  \nonumber &=& a_1\int^{a_0}_{a_1}\frac{da}{a^2H_0\sqrt{\Omega_\Lambda+\Omega_M(\frac{a_0}{a})^3+\Omega_R(\frac{a_0}{a})^4}} \\
   &\simeq& \frac{3.2}{H_0}\frac{a_1}{a_0}\simeq\frac{3.2}{H_0}\Omega_R^{1/4}\left(\frac{H_0}{H_1}\right)^{1/2},
\end{eqnarray}
where we have used approximations $\Omega\simeq1$ in the first sim equality because $\rho(t)<\rho_1\ll\rho_c$ when $t>t_1$, and in the second and third sim equality, $\frac{a_1}{a_0}\sim 0$.
So, for solving the horizon problem, we need $d_H(t_1)>\frac{3.2}{H_0}\Omega_R^{1/4}\left(\frac{H_0}{H_1}\right)^{1/2}$, i.e. $\ln\left(\sqrt{H_1}d_H(t_1)\right)>\text{ln}\left(3.2\left(\frac{\Omega_R}{H_0^2}\right)^{1/4}\right)\simeq69$, where we have used the Planck units ($G=\hbar=c=1$). It may be noted that the number 69 is indeed related to 68 $e$-foldings in the literature (e.g.,\cite{Weinberg}), which is needed for an exponential inflation to solve the horizon problem at Planck energy density (i.e., choose $\rho_1=\rho_P$), but which is unnecessary because at the end of inflation, the density $\rho_1\ll\rho_P$. Here we get the number because of choosing the Planck units; however, interestingly, if the $e$-foldings is defined to contain not only inflation period but also the period before inflation, then the results of the numerical calculations in Secs.VI and VII will show that 68 $e$-foldings is precisely the least number for solving the horizon problem.

\section{\label{sec:level4} The measure of space of solutions from Liouville measure }
Before considering some concrete models, we need to define a natural measure to measure the probability of inflation. The Liouville measure, first used in \cite{Gibbons1}, is a candidate scheme. In the canonical framework as in Sec.II, we have the symplectic two-form $\omega=\frac{1}{4\pi G\gamma\sqrt{\Delta}}d{\cal{V}}\wedge db+dp_\phi\wedge d\phi$. For this constraint system, physical points are on the constraint surface $\bar\Gamma :C_{\text{eff}}=0$. By restricting $\omega$ onto $\bar\Gamma$, one gets $\omega|_{\bar\Gamma}=\frac{1}{4\pi G\gamma\sqrt{\Delta}}d{\cal{V}}\wedge db\pm d\left(\cal{V}\sqrt{\frac{6}{\kappa\gamma^2\Delta}\sin^2(b)-2V(\phi)}\right)\wedge d\phi$. Because $b$ decreases with time, one can view $b$ as an internal time; thus, for every fixed $b$, $\omega|_{\bar\Gamma,b=const}=\pm \sqrt{\frac{6}{\kappa\gamma^2\Delta}\sin^2(b)-2V(\phi)}d{\cal{V}}\wedge d\phi$. Thus, at any fixed $b$, $d\mu_b=\sqrt{\frac{6}{\kappa\gamma^2\Delta}\sin^2(b)-2V(\phi)}d{\cal{V}}\wedge d\phi$ can be used as a natural measure for the space of solutions of the constraint system, for its invariance with time, $b$. However, some subtle problems appear: on the one hand, the total volume of space of solutions would be infinity under this measure. This is because the range of variable $\cal{V}$ is infinite [It is worth mentioning that the range of $\phi$ may also be infinite for some special potential $V(\phi)$, but we will not consider this case in this paper]. On the other hand, the volume of element cell $\cal{V}$ or the scale factor $a$ is a gauge quantity. One can cure these by a trick used in \cite{Ashtekar}: by introducing a cutoff $[\frac{1}{\cal{V}^*},\cal{V}^*]$ of $\cal{V}$,  and integrating out this variable, one can get a new measure of the solutions' space
\begin{eqnarray}
 \nonumber
d\tilde{\mu}_b &\propto &\frac{\int_{\frac 1{\cal{V}_{*}}}^{\cal{V}%
_{*}}d\mu _b}{\int_\phi \int_{\frac 1{\cal{V}_{*}}}^{\cal{V}%
_{*}}d\mu _b} \\
&\propto &\sqrt{\frac 6{\kappa \gamma ^2\Delta }\sin ^2(b)-2V(\phi )}d\phi
\equiv |\dot{\phi}|d\phi ,
\end{eqnarray}
which is independent of the cutoff $\cal{V}^*$. But as the authors of Ref.\cite{Corichi} said, such a measure depends on the choice of time $b$ when integrating out volume. We can also see this thing by comparing the probability of two small regions, for instance, $[\phi_1,\phi_1+\delta\phi_1]$ and $[\phi_2,\phi_2+\delta\phi_2]$ at time $b$. Suppose they become $[\tilde{\phi}_1,\tilde{\phi}_1+\delta\tilde{\phi}_1]$ and $[\tilde{\phi}_2,\tilde{\phi}_2+\delta\tilde{\phi}_2]$ at time $\tilde b$. In order to use the invariance of the Liouville measure, introduce volume variable $\cal{V}$ for the two regions $[{\cal{V}}_1,{\cal{V}}_1+\Delta{\cal{V}}_1]$ and $[{\cal{V}}_2,{\cal{V}}_2+\Delta{\cal{V}}_2]$ at time $b$, satisfying $\Delta{\cal{V}}_1=\Delta{\cal{V}}_2$. Suppose the cell volumes expand by some factors $e^{3N_1}$ and $e^{3N_2}$ at time $\tilde{b}$, then  we have the simple relations $\Delta\tilde{{\cal{V}}}_1=e^{3N_1}\Delta{\cal{V}}_1$ and $\Delta\tilde{\cal{V}}_2=e^{3N_2}\Delta{\cal{V}}_2$, because the equation of $\ln\cal{V}$ does not depend on variable $\cal{V}$: $\frac{d\ln\cal{V}}{db}=-\rho_c\frac{\sin(2b)}{\dot\phi^2}$. Then,
\begin{eqnarray}\nonumber
  \frac{Pro|_b([\phi_1,\phi_1+\delta \phi_1])}{Pro|_b([\phi_2,\phi_2+\delta \phi_2])} &=& \frac{|\dot\phi_1|\delta\phi_1}{|\dot\phi_2|\delta\phi_2} =\frac{|\dot\phi_1|\delta\phi_1\Delta {\cal{V}}_1}{|\dot\phi_2|\delta\phi_2\Delta {\cal{V}}_2}\\
  =\frac{|\dot{\tilde\phi}_1|\delta{\tilde\phi}_1\Delta \tilde{\cal{V}}_1}{|\dot{\tilde\phi}_2|\delta{\tilde\phi}_2\Delta \tilde{\cal{V}}_2} &=& \frac{Pro|_{\tilde{b}}([\tilde{\phi}_1,\tilde{\phi}_1+\delta \tilde{\phi}_1])}{Pro|_{\tilde{b}}([\tilde{\phi}_2,\tilde{\phi}_2+\delta \tilde{\phi}_2])}\frac{e^{3N_1}}{e^{3N_2}},
\end{eqnarray}
where we have used the property that $|\dot{\phi}|\delta\phi\Delta\cal{V}$ is invariant with time in the third equality. Thus, if one who gets the measure at time $b$  finds equal probability of region 1 and region 2, another person who gets the measure at time $\tilde b$ will find region 1 is more impossible than region 2 if $N_1>N_2$. So when to integrate out volume is crucial, especially for a system having both inflation solutions and noninflation solutions. Our view, in this paper, is to take the time integrating out volumes at the starting time of the Universe, and LQC has such a starting time, when the density $\rho$ is $\rho_c/2$, or $\sin^2b=1/2$. So we will use
\begin{equation}\label{measure}
  d\mu\propto \sqrt{\rho_c/2-V(\phi_0)} d\phi_0.
\end{equation}
An interesting thing is that the measure of classical theory is also the above formula, if one takes the initial condition at $\rho=\rho_c/2$ in classical theory.
In the next two sections we will consider two inflation models; one is the quadratic inflation, the most simple model although it has been disfavored for recent observation, and the other is the natural inflation.

\section{\label{sec:level5}Quadratic inflation model}

\begin{figure}[h]
\includegraphics[clip,width=0.35\textwidth]{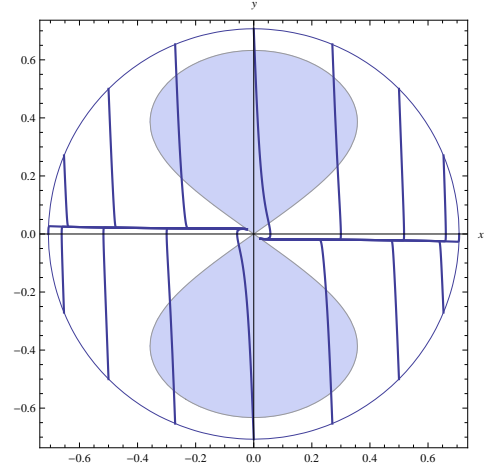}
\caption{Phase portrait for quadratic inflation, where $x\equiv \frac{m\phi}{\sqrt{2\rho_c}}$ and $y\equiv \frac{\dot\phi}{\sqrt{2\rho_c}}$. The shaded area denotes decelerating region: $\ddot{a}<0$, and inflation ends at $\ddot{a}=0$. We use mass $m=1.06\times10^{-1}.$   }
\label{Fig.1}
\end{figure}

As a simplest example studied in many references, we first consider a massive scalar field $\phi$ with mass $m$, i.e. a scalar field with quadratic potential
\begin{equation}\label{potential1}
  V(\phi)=\frac{1}{2}m^2\phi^2.
\end{equation}
We choose the mass of the scalar field $m=1.06\times 10^{-6}m_P$, which gives the wanted amplitude of curvature perturbations $A_S=10^{-10}e^{3.062}$ and the spectral index $n_S=0.97$, but the unwanted tensor-to-scalar ratio, $r=0.13$.
Following from \cite{Linsefors}, define:
\begin{equation}\label{variables}
  x:=\frac{m\phi}{\sqrt{2\rho_c}} ~~\text{and}~~ y:=\frac{\dot\phi}{\sqrt{2\rho_c}},
\end{equation}
so the density becomes $\rho=\rho_c(x^2+y^2)$. For the numerical calculation, define the conformal causal distance $D_H(t):=\int_0^t\frac{\sqrt{\Omega}}{a}dt$, then the equations of $x$, $y$, scale factor $a$, and $D_H$ are (in Planck units)
\begin{equation}
\left\{
\begin{array}{l}
\dot x=my, \\
\dot y=-mx-3Hy, \\
\frac{\dot a}{a}=H\equiv \sqrt{\frac{8 \pi\times 0.41}{3}}\sqrt{(x^2+y^2)(1-x^2-y^2)}, \\
\dot{D}_H=\frac{\sqrt{1-2(x^2+y^2)}}{a}.
\end{array}
\right.   \label{eom}
\end{equation}
The initial values are
\begin{equation}
\left\{
\begin{array}{l}
x(0)=\cos(\theta)/\sqrt{2}, ~~~\theta\in[0,2\pi) \\
y(0)=\sin(\theta)/\sqrt{2}, \\
a(0)=1, \\
D_H(0)=0.
\end{array}
\right.   \label{iv}
\end{equation}
Figure \ref{Fig.1} shows the phase space trajectories for quadratic inflation, for better viewing, we've chosen $m=1.06\times10^{-1}m_P$ instead of $m=1.06\times10^{-6}m_P$. From the picture, we can find every initial point on $\rho=\rho_c/2$ quickly enters in slow roll solution region.

In order to determine which solutions do not have the horizon problem, i.e., satisfy $\ln(\sqrt{H_1}d_H(t_1))>69$, one can use numerical methods, whose results have been shown in Table \ref{Table 1}. In the following, we give the slow roll approximation which is shown to be an excellent approximation.

If a solution has entered into the slow roll region where the trajectories satisfy $|y|\ll|x|$ and $0=-mx-3Hy$, then the modified Friedmann equation becomes $H^2=\frac{\kappa}{3}V(1-V/\rho_c)=\frac{\kappa\rho_c}{3} x^2(1-x^2)$ . Then the slow roll solution is $y=-\frac{mx}{3H}=-\frac{m \text{sgn}(x)}{\sqrt{3\kappa\rho_c}\sqrt{1-x^2}}\sim  -2.5\times10^{-7}\text{sgn}(x)$ which is in the second or fourth quadrant. When the density is much lower than Planck density and the potential slow roll parameter $\epsilon_V:=\frac{1}{16\pi G}\left(\frac{V'}{V}\right)^2=\frac{m_P^2}{4\pi \phi^2}$ equals to 1, inflation ends and $\phi_{\text{end}}=\pm \frac{m_P}{\sqrt{4\pi}}$.

The $e$-foldings during inflation can be approximated by
\begin{eqnarray}\label{e-foldings}
  \nonumber &N& \simeq\int_{t_\star}^{t_{\text{end}}}Hdt=\int_{\phi_\star}^{\phi_{\text{end}}}\frac{H}{\dot\phi}d\phi=\int_{x_\star}^{x_{\text{end}}}\frac{H}{y}dx \\
  &\simeq&\frac{4\pi\rho_c}{m_P^2m^2}x_\star^2\left(1-\frac{x_\star^2}{2}\right)-\frac{1}{2},
   \end{eqnarray}
where $t_\star$ denotes the starting time of inflation.
We need to associate the starting point $(x_0,y_0)$ with the inflation starting point $x_\star$. For doing this, consider $\frac{dx}{dy}=\frac{\dot x}{\dot y}=-\frac{1}{\frac{x}{y}+\frac{3H}{m}}$. When $|y|\gtrsim 10^{-6}$, we get $\frac{3H}{m}>1$ and $\frac{3H}{m}>|\frac{x}{y}|$, so the kinetic energy decays more quickly than the potential, and then the Universe will enter into the slow roll inflation region

(a) If the slow roll part is in the fourth quadrant ($y<0$), we use the approximation:
$x_1:= x_0+\int_{10^{-6}}^{y_0}\frac{dy}{\frac{x_0}{y}+\frac{3H|_{x_0}}{m}}$, and
\begin{equation}
x_\star \simeq x_1+\int_0 ^{10^{-6}}\frac{dy}{\frac{x_1}{y}+\frac{3H|_{x_1}}{m}};
\end{equation}

(b) If the slow roll part is in the second quadrant ($y>0$), we use the approximation:
$x_1:= x_0+\int_{10^{-6}}^{y_0}\frac{dy}{\frac{x_0}{y}+\frac{3H|_{x_0}}{m}}$, and
\begin{equation}
x_\star \simeq x_1+\int_{3\times 10^{-7}} ^{10^{-6}}\frac{dy}{\frac{x_1}{y}+\frac{3H|_{x_1}}{m}}.
\end{equation}

\begin{widetext}
\begin{center}
\begin{table}[h]
\caption{Quadratic inflation. We only present results in $[0,\pi]$, for the symmetry: $(x,y,a,D_H)$ and $(-x,-y,a,D_H)$ are both solutions of Eqs. (\ref{eom}) and (\ref{iv}). In the fourth column, $N$ only contains the inflation period and is approximated by slow roll approximations. }\label{Table 1}
\begin{ruledtabular}
\begin{tabular}[c]{ccccc}
  $\theta$ & $\ln(\sqrt{H_1}a_1D_{H1})$ & $N=a_1/a(0)$ & $N$(inflation) \\
  \hline
  0 & $1.71954\times10^{12}$ & $1.71954\times10^{12}$ & $1.71954\times10^{12}$ \\
  \hline
  $\pi/4$ & $1.00307\times10^{12}$ & $1.00307\times10^{12}$ & $1.00307\times10^{12}$ \\
  \hline
  $1.5707946$ & $68.9633$ & $67.9929$ & $65.3867$ \\
  \hline
  $\pi/2$ & $36.2368$ & $35.0089$ & $31.573$ \\
  \hline
  $1.57080507$ & $68.9654$ & $68.0229$ & $64.5201$ \\
  \hline
  $3\pi/4$ & $1.00307\times10^{12}$ & $1.00307\times10^{12}$ & $1.00307\times10^{12}$\\
\end{tabular}
\end{ruledtabular}
\end{table}
\end{center}
\end{widetext}

By these approximations, one can find $N(\theta=\pi/2)=32$, $N(1.5707945)=68$, $N(1.57080)= -0.4$, and $N(1.5708052)=68$. So when $\theta\in[1.5707945,1.5708052]$, the $e$-foldings is smaller than 68. The approximating results are also presented in Table 1, from which we can see the approximation is excellent. The measure in Eq.(\ref{measure}) becomes $d\mu\propto\sqrt{\frac{\rho_c}{2}-\rho_c x_0^2}dx_0\propto \sin^2(\theta)d\theta$, and from the result in Table \ref{Table 1}, we get the probability of enough inflation is $1-Pro(\theta\in[1.5707946,1.57080507])\simeq 99.9989\%$. This result is not quite different from previous works in Ref.\cite{Ashtekar} in the framework in LQC and even similar to the classical theory\cite{Belinsky}. It is expected and one can be see this from two aspects:
on the one hand, we have said the measure of the classical model is same with LQC, if one uses the initial
time at the same density; one the other hand, we showed near Eq.(12) that for matters with the density formula in Eq.(11),
if an initial value can solve the horizon problem in LQC, then it must solve the problem in GR too. We can
guess that this conclusion is also almost true for other matters, then because for $\phi^2$ model LQC gives near
100\% probability, GR would also give the result. This guess is reasonable because the probability
calculation only depends on those solutions which give just enough of the wanted $e$-foldings, and
are dominated by kinetic energy at the beginning time (i.e., $P\simeq \rho$), and then becomes classical, which means the difference is not large.
We want to say that the reason to give the initial value at $\rho=\rho_c/2$ is from LQC's perturbation, although
the results are not quite different.

\section{\label{sec:level6}Natural inflation model}

For the natural or cosine inflation model \cite{Adams,Freese}, the inflaton $\phi$ is a pseudo-Nambu-Goldstone boson $\theta=\phi/f$ with a global shift symmetry broken at scale $f$, and the potential has the form of
\begin{equation}\label{potential2 }
  V(\phi)=\Lambda^4[1-\cos(\phi/f)],
\end{equation}
where $\Lambda$ denotes the energy scale where the global symmetry is broken. The probability of natural inflation in classical theory has been investigated via a different measure in \cite{Remmen}; in this section, we will use LQC to study this model. To fit with recent observation \cite{Planck}, we choose the parameters $\Lambda=1.24\times 10^{-3}m_P$ and $f=7M_P=1.39m_P$, where $M_P=1/\sqrt{8\pi G}$ is the reduced Planck mass, which gives the amplitude of curvature perturbations $A_S=10^{-10}e^{3.061}$, the spectral index $n_S=0.96$, and the tensor-to-scalar ratio $r=0.07$. Note that for the small value of $\Lambda\ll m_P$, every solution's density must be much lower than Planck energy when it enters in an inflation region where the potential energy dominates energy density; thus, quantum geometry's corrections will not be important for this model when solutions enter into slow roll region. Define dimensionless variables:
\begin{equation}\label{variables2}
  x:=\frac{\phi}{2f} ~~\text{and}~~ y:=\frac{\dot\phi}{\sqrt{2\rho_c}}.
\end{equation}
Then the density is $\rho=\rho_c\left(\frac{2\Lambda^4}{\rho_c}\sin^2(x)+y^2\right)\equiv \rho_c(\xi^2+y^2)$, where we have defined $\xi:=\sqrt{\frac{2}{\rho_c}}\Lambda^2\sin(x)\sim 3.4\times10^{-6}\sin(x)$.
The equations of motion are (in Planck units)
\begin{equation}
\left\{
\begin{array}{l}
\dot x=\frac{\sqrt{0.41/2}}{f}y, \\
\dot y=-3Hy-\frac{\Lambda^4}{2f\sqrt{\rho_c/2}}\sin(2x), \\
\frac{\dot a}{a}=H\equiv \sqrt{8 \pi\times 0.41/3}\sqrt{(\xi^2+y^2)(1-\xi^2-y^2)}, \\
\dot{D}_H=\sqrt{1-2(\xi^2+y^2)}/a.
\end{array}
\right.   \label{eom2}
\end{equation}
Initial values are
\begin{equation}
\left\{
\begin{array}{l}
x(0)=x_0\in[-\frac{\pi}{2},\frac{\pi}{2}], \\
y(0)=\pm \frac{1}{\sqrt{2}}\sqrt{1-\frac{4\Lambda^4}{\rho_c}\sin^2{x_0}}, \\
a(0)=1, \\
D_H(0)=0.
\end{array}
\right.   \label{iv2}
\end{equation}

\begin{figure}[h]
\includegraphics[clip,width=0.35\textwidth]{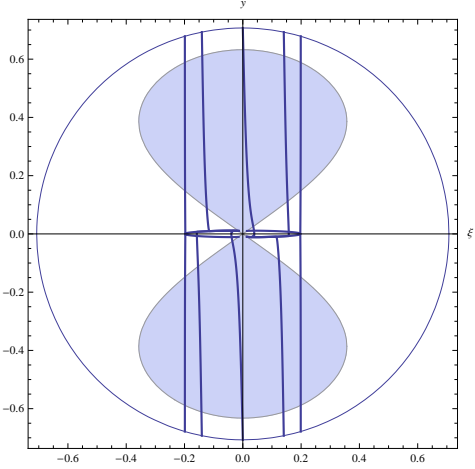}
\caption{Phase portrait for quadratic inflation, where $\xi\equiv \sqrt{\frac{2}{\rho_c}}\Lambda^2\sin(\phi/f)$ and $y\equiv \frac{\dot\phi}{\sqrt{2\rho_c}}$ for natural inflation. We use the parameter $\Lambda=0.3m_P$. The shaded area denotes decelerating region: $\ddot{a}<0$. There are four slow roll inflation attractors in this model. }
\label{Fig.2}
\end{figure}

Figure \ref{Fig.2} shows some phase space trajectories for natural inflation, from which we know that there exist four inflation attractors in each quadrant, which can also be obtained by slow roll approximation as follows.

The slow roll parameter is $\epsilon_V=\frac{1}{16\pi G}\left(\frac{V'}{V}\right)^2=\frac{1}{16\pi}\left(\frac{m_P}{f}\right)^2\frac{1}{\tan^2(x)}$, so the ending $x$ satisfies $|\tan(x_{\text{end}})|=\frac{M_P}{\sqrt{2}f}=\frac{1}{7\sqrt{2}}$.
A general slow roll solution is $y=\frac{1}{\sqrt{2\rho_c}}\frac{-V'}{3H}=-\frac{\text{sgn}(\xi)}{\sqrt{24\pi\rho_c}}\frac{\Lambda^2}{f}\cos(x)$, which contains four slow roll inflation solutions that agreed with Fig.\ref{Fig.2}.
The $e$-foldings during inflation can be approximated by
\begin{eqnarray}\label{e-foldings2}
 \nonumber N& \simeq&\int_{\phi_\star}^{\phi_{\text{end}}}\frac{H}{\dot\phi}d\phi=\int^{\phi_\star}_{\phi_{\text{end}}}8\pi G\frac{V}{V'}(1-\frac{V}{\rho_c})d\phi
   \\
   &\simeq&\int^{\phi_\star}_{\phi_{\text{end}}}8\pi G\frac{V}{V'}=98 \ln\left(\frac{\sqrt{98/99}}{|\cos(x_\star)|}\right).
\end{eqnarray}
From this formula, one can find when $x_\star=\frac{\pi}{2}+k\pi$, $k\in\mathbb{Z}$, $N\rightarrow \infty$. And if one wants $N>68$ (for simplicity, we use 68 $e$-foldings), he needs to require $x_\star\in(\hat x, \pi-\hat x)\cup (\pi+\hat x, 2\pi-\hat x)\cup...$, where $\hat x:=\arccos\left(\sqrt{\frac{98}{99}}\exp(-\frac{68}{98})\right)\simeq1.05054$.

Similar to the quadratic model, the relation between initial points and inflation starting points is needed when considering the probability of inflation. Because of the shift symmetry, initial values are restricted in region $x_0\in[-\frac{\pi}{2},\frac{\pi}{2}]$, however, $x_\star$ can be out of this region.

For the symmetry of the space of solutions, we only need to consider the cases of $y(0)>0$.

Firstly, let us compare the variation rate of kinetic energy with potential energy:
\begin{equation}
\frac{\dot y}{\dot \xi}=-\frac{3f}{\Lambda^2}\frac{H}{\cos(x)}-\frac{\xi}{y}.
\end{equation}

(a) In the right hand side of the equation, if we require $|\frac{3f}{\Lambda^2}\frac{H}{\cos(x)}|>|\frac{\xi}{y}|$, then $\frac{6f}{\Lambda^4}\sqrt{\frac{\rho_c}{2}}Hy>|\sin(2x)|$, where $\frac{6f}{\Lambda^4}\sqrt{\frac{\rho_c}{2}}H\sim 3\times10^{12}\sqrt{\xi^2+y^2}$, which means $|y|>5\times10^{-7}$ will satisfy the requirement. Furthermore, if $x\rightarrow\frac{\pi}{2}$, any $y$ will satisfy this requirement;

(b) The term $|\frac{3f}{\Lambda^2}\frac{H}{\cos(x)}|\sim5\times10^6\frac{\sqrt{\xi^2+y^2}}{|\cos(x)|}>10$, when $|y|>2\times10^{-6}$. So in this region, kinetic energy decays quickly, and then the system enters into slow roll region.

Secondly, to find out $x_\star$'s dependence on $x_0$, let us consider the following equation;
\begin{eqnarray}
\frac{dx}{dy} &=&\frac 1{d\xi /dx}(\frac{\dot{y}}{\dot{\xi}})^{-1}=-\frac 1{%
3f\sqrt{2/\rho _c}H+\frac{\Lambda ^4}{\rho _c}\frac{\sin (2x)}y}  \nonumber \\
&\equiv &-F(x,y).
\end{eqnarray}

(a) If the slow roll part is in $y>0$, one can use:$x_1:= x_0+\int_{2\times10^{-6}}^{y_0}F(x_0,y)dy$; then we have
\begin{equation}
x_\star \simeq x_1+\int_{5\times10^{-7}} ^{2\times10^{-6}}F(x_1,y)dy;
\end{equation}

(b) If the slow roll part is in $y<0$, one can use $x_1:= x_0+\int_{2\times10^{-6}}^{y_0}F(x_0,y)dy$, then we have
\begin{equation}
x_\star \simeq x_1+\int_0 ^{2\times10^{-6}}F(x_1,y)dy.
\end{equation}

Some special points are the following: $x_\star(x_0=0.2247)\simeq\hat x$, $x_\star(0.8053)\simeq\frac{\pi}{2}$, $x_\star(1.3267)\simeq\pi-\hat x$, and $x_\star(-\frac{\pi}{2})\simeq-0.8362$.  So only when $x_0\in[0.2247,1.3267]$, we have $N>68$.

The measure in Eq.(\ref{measure}) in this model is
\begin{eqnarray}
d\mu\propto\sqrt{\frac{\rho_c}{2}-2\Lambda^4\sin^2(x_0)}dx_0\propto\sqrt{1-\frac{4\Lambda^4}{\rho_c}\sin^2(x_0)}dx_0. \nonumber \\
\end{eqnarray}
So the probability of those solutions which have enough $e$-foldings is $\frac{\int_{0.2247}^{1.3267}d\mu}{\int_{-\frac{\pi}{2}}^{\frac{\pi}{2}}d\mu}\simeq35.0777\%$.
The results of numerical methods of calculating horizon are exhibited in Table \ref{Table 2}, where the results of slow roll approximation are also exhibited. From the calculations, we find that initial values in the region $x_0\in[0.241,1.347]$ do not have the horizon problem, whose probability is $\frac{\int_{0.241}^{1.347}d\mu}{\int_{-\frac{\pi}{2}}^{\frac{\pi}{2}}d\mu}\simeq35.2051\%$.

\begin{widetext}
\begin{center}
\begin{table}[h]
\caption{Natural inflation. We only present results in $y_0>0$, for the symmetry: $(x,y,a,D_H)$ and $(-x,-y,a,D_H)$ are both solutions of Eqs.(\ref{eom2}) and (\ref{iv2}). The first column denotes the quadrant where the inflation attractors are. From the table, there exists a point $x_0$ near 0.8036, such that $N\rightarrow \infty$.}\label{Table 2}
\begin{ruledtabular}
\begin{tabular}[c]{ccccc}
  Attractor&$x_0 ( y_0>0)$ & $\ln(\sqrt{H_1}a_1D_{H1})$ & $N=a_1/a(0)$ & $N$(inflation) \\
  \hline
  $I$&$\pi/2$ & $39.9852$ & $38.7343$ & $32.2118$ \\
  \hline
  $I$&$1.347$ & $69.0734$ & $67.9437$ & $64.2177$ \\
  \hline
  $I$&$0.8036181717$ & $2270.63$ & $2269.60$ & $651.831$ \\
  \hline
  $IV$&$0.8036181716$ & $2411.48$ & $2410.45$ & $651.831$ \\
  \hline
  $IV$&$0.241$ & $69.0219$ & $67.8726$ & $62.3837$ \\
  \hline
  $IV$&$0$ & $40.3895$ & $39.1099$ & $34.3456$ \\
  \hline
  $III$&$-\pi/2$ & $39.9852$ & $38.7343$ & $34.5361$\\
\end{tabular}
\end{ruledtabular}
\end{table}
\end{center}
\end{widetext}

\section{\label{sec:level7} conclusions}
Inflation can solve many cosmological problems; moreover, it can also give a natural explanation of the structure formation. More and more precise observations such as the power spectra of perturbations have also been observed which could constrain inflation models strictly. On the other hand, the genericness for an inflation model is also an important indirect constraint for inflation models.

Anomaly-free perturbations of LQC reveal some new aspects of loop quantum cosmology. The constraint algebra of LQC is deformed, and LQC also has a starting time similar to the big bang theory of cosmology; however, there does not exist any singularity in LQC. The starting point of time also helps to choose a natural measure. In this paper, we used the effective causal structure of LQC, and considered the horizon problem in LQC and its resolution by inflation. The probabilities of two inflation models are calculated. We find that for the quadratic inflation the probability of a sufficient inflation is close to $1$, while for the natural inflation the probability is about $35\%$ smaller than the quadratic model. Of course the probability for the natural inflation is not large, but it is still much more probable than the classical cosmological fine-tuning initial value. It should be noted that the conclusions in this paper almost agree with previous works \cite{Ashtekar,Ashtekar2,Linsefors}, even agree with the classical theory \cite{Belinsky,Kofman,Remmen}. But we view this work as a
self-contained work in the framework of loop quantum cosmology, such as the horizon problem and how many $e$-foldings needed to solve the horizon problem in loop quantum cosmology. Although, only two models are considered in this paper, we could conclude safely that, for different models there could be quite different probabilities for enough inflation.

It should be noted that we did not consider the anisotropies, which are interesting and important problems that have been considered in Refs.\cite{Gupt,Linsefors2014a,Linsefors2014b,Fujio}. In Refs.\cite{Gupt,Linsefors2014a,Fujio}, the authors proposed initial conditions at bounce while the authors of Ref.\cite{Linsefors2014b} considered at remote past before bounce. They showed that the shear term will decrease the possibility of inflation for the latter, while the former faces a problem that there exists infinite regions of solutions never reach the classical behavior which is severer. From the view in the present work, taking initial conditions near (after) the bounce may be a natural one, because the perturbation of the Friedmann-Robertson-Walker model should be some limit of the perturbation of Bianchi-I model. But whether these initial conditions (or some more precise conditions) can exclude those regions where the solutions never reach the classical behavior is beyond this work and deserves future research. If this problem could be solved, then from the work in Refs.\cite{Gupt,Fujio}, isotropic universes or inflation seem to be favored. We hope the anomaly-free perturbations of the Bianchi-I model could be derived in the future and may help to answer this question.

\acknowledgments This work was supported by the National Natural Science Foundation of China (Grants No. 11175019 and No. 11235003).

\end{document}